# Technoeconomic Analysis of Thermal Energy Grid Storage Using Graphite and Tin

Colin C. Kelsall[1], Kyle Buznitsky[1], Asegun Henry[1]
[1]Department of Mechanical Engineering
Massachusetts Institute of Technology, Cambridge MA, 02139

Address correspondence to: ase@mit.edu

**Introduction:**

Energy storage is needed to enable dispatchable renewable energy supply and thereby full decarbonization of the grid. However, this can only occur with drastic cost reductions compared to current battery technology, with predicted targets for the cost per unit energy (CPE) below $20/kWh [1–3]. Notably, for full decarbonization, long duration storage up to 100 hrs will be needed at such low costs, and prior analyses have shown that in such high renewable penetration scenarios, CPE is more critical than other parameters such as round trip efficiency (RTE) or cost per unit power (CPP) when comparing the costs of different technologies. Here, we introduce an electricity storage concept that stores electricity as sensible heat in graphite storage blocks and uses multi-junction thermophotovoltaics (TPV) as a heat engine to convert it back to electricity on demand. This design is an outgrowth of the system proposed by Amy *et al.* in 2019,[4] which has been modified here to use a solid graphite medium and molten tin as a heat transfer fluid rather than silicon as both. The reason for this is two-fold: (1) the CPE of graphite is almost 10X lower than that of silicon, which derives from the lower cost per unit mass (i.e., $0.5/kg vs. $1.5/kg) and the higher heat capacity per unit mass (2000 J kg$^{-1}$ K$^{-1}$ vs. 950 J kg$^{-1}$ K$^{-1}$); and (2) the melting point of tin and solubility of tin in graphite are much lower than that of silicon, which lessens the number of issues that have to overcome along the research and development (R&D) pathway. The usage of graphite also eliminates the need for a second tank, but the main disadvantage of using a solid medium is that one cannot easily provide a steady discharge rate, as the power output from the storage will change with time, as the graphite cools during discharge. Thus, the objective of this work is to examine how these changes in the system design effect the overall technoeconomics. The technoeconomic analysis presented in Amy's work was repeated here (i.e., using the same approach), but updated and modified to reflect the design changes, and this document provides a summary of this analysis.

**Proposed System Design:**

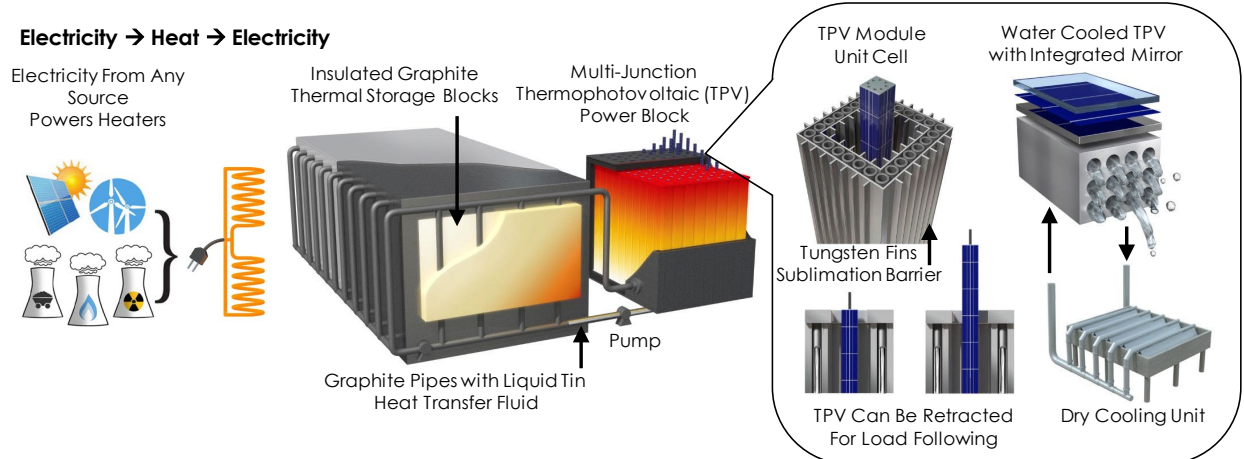

**Figure 1:** Proposed TEGS concept using graphite storage blocks, a molten tin heat transfer fluid, and a multijunction TPV heat engine.

The envisaged system is illustrated in Fig. 1, whereby excess electricity from the grid is first used to power resistive heating elements (i.e., made of either graphite or tungsten), which then convert the electricity to extremely high temperature heat (~ 2500C). The energy is then transferred to graphite pipes via thermal radiation, and the pipes carry liquid tin is, which is used as a heat transfer fluid (HTF). Nominally, just like in the prior work of Amy et al., the tin is envisaged to be heated from 1900C up to 2400C, thereby converting the energy input into sensible heat in the tin, by raising its enthalpy. The tin is pumped through the piping continuously, and is then routed to the storage unit, which contains large graphite blocks. As the 2400C tin is pumped through pipes that run through the graphite blocks, the tin heats the graphite blocks up from 1900C to 2400C via thermal radiation, which correspondingly causes the tin to cool back to 1900C. The tin is then pumped back through the resistance heaters to be reheated, and this process, which constitutes the charging process, continues until the graphite storage is fully heated to the peak temperature.

Since the graphite storage unit is large, on the order of 1000 $m^3$, its thermal mass is sufficiently large, that it can retain the energy used to charge it for long periods of time (e.g., multiple days or even > 1 week) with minimal i.e., < 10% loss of the energy stored – note that the heat loss design point in prior work by Amy et al., was 1% loss per day, and this is significant/important design choice. Thus, the energy is stored as sensible heat in the graphite until electricity is needed again. When electricity is desired, the system is discharged by pumping liquid tin through the graphite storage unit, which heats it to the peak temperature 2400C, after which it is routed to the power block. The power block consists of an array of graphite pipes that form vertically oriented unit cells. Each unit cell of piping creates a square or rectangular cavity that is lined with tungsten foil, which is used as a diffusion barrier to prevent graphite deposition onto the MPV cells. Inside each cavity, a water/oil cooled heat sink that is covered in MPV cells can be lowered into the unit cell cavity. This causes the MPV cells to be illuminated with the light emitted by the tungsten foil, which is in turn heated by the light emitted by the graphite piping carrying the tin. This net transfer of energy converts a large fraction ≥ 50% of the energy to electricity, which in turn causes the tin's temperature to decrease back to 1900C, and the remaining waste heat is removed by the coolant running through the heat sink. The coolant subsequently dissipates the waste heat to the environment via a dry cooling unit, while the tin is pumped back to the graphite storage unit to be reheated by the stored sensible heat. In this way the entire system serves effectively as a rechargeable grid scale thermal battery that can store energy cheaply and supply it to the grid on demand, with an estimated RTE of ~ 50%. It should be noted here that as concluded by Amy et al., the RTE is almost fully determined by the MPV conversion efficiency, and since the conversion is predominated by infrared light, the MPV cells can also be termed thermophotovoltaic (TPV) cells.

**Technoeconomic Analysis:**

To properly capture the range of useful implementations of energy storage, the predicted costs have been split into CPP in the units of dollars per watt-electric of the power block, and CPE in the units of dollars per kWh-electric of storage capacity. Four potential systems are considered as well: a 100MW-10 hour (1 GWh-e) system, a 100 MW-4 hour (400 MWh-e) system, a 10 MW-10 hour (100 MWh-e) system, and a 10 MW-4 hour (40 MWh-e) system. This technoeconomic model replicates the same methodology and references for pricing as Amy *et al.*[4], with added cost components for the heat transfer fluid and increased quantities of graphite. Tin was chosen as the heat transfer fluid based on previous work showing that it can be pumped and contained within an all graphite infrastructure.[5] The cost of tin was taken as $17/kg based on a review of bulk tin prices. The material prices used for this analysis are shown in Table 1. This analysis does not consider site-specific development and balance-of-plant costs.

**Table 1:** Bulk material prices[4,6]

| Material | Density (kg/m$^3$) | Cost ($/kg)[4,6] |
|---|---|---|
| Extruded Graphite | 1700 | 0.5 |
| Graphite Insulation | 24 | 540 |
| Aluminum Silicate Insulation | 100 | 4.0 |
| Fiberglass Insulation | 12 | 7.1 |
| Tungsten Foil | 19000 | 350 |

In general, the cost effectiveness/profitability of an energy storage system is determined by three main quantities: (1) the initial one-time total capital expenditure to create and install the system (CAPEX), (2) the recurring costs associated with operating the system (OPEX), and (3) its lifetime i.e., the number of charge/discharge cycles it can endure, before requiring replacement. Based on prior analysis, the predicted OPEX for the proposed system shown in Fig. 1 is very low ($X/kW-yr), and the lifetime is anticipated to be > 30 years, since there is no known physical degradation mechanism within the system that would require replacement of components on a time scale less than 100 yrs. Therefore, the focus in this analysis is on the main cost component that would determine its cost effectiveness relative to other competing options, which is its CAPEX. The CAPEX here, denoted by K ($/kWh) is given by a combination of the CPE ($/kWh), CPP ($/kW), discharge duration t (hours) as follows: K = (CPE + CPP/t). With these inputs, one can estimate the total CAPEX for a given system configuration, and the major changes that occur in this work, as compared to the prior work of Amy et al., are associated with CPE, since the rest of the system is essentially unchanged.

In Fig. 2, the calculated CPE is shown, which represents the cost of one kWh-e of storage capacity,

independently of all the components that scale with the charge/discharge rate of the system. This results show how changing the size and storage duration affects the system cost. Here, it becomes clear that the graphite insulation, used to retain the heat is the predominant cost that changes with size. This is due to the change in surface area to volume ratio, and heavily favors larger system sizes, where the fixed insulation thickness (i.e., skin thickness) comprises a smaller fraction of the total volume. It is also important to note here that the heat loss rate can be adjusted to affect this cost significantly. For the results in Fig. 2, the heat loss rate was limited to 1% of the energy stored being lost each day. If, for example, one was building a smaller system that was expected to discharge every day, one might be able to tolerate a higher heat loss rate or > 2%, which could significantly reduce the costs associated with smaller systems. Furthermore, when systems are made very large, one could potentially afford to reduce the heat losses below 1%. Nonetheless, the results in Fig. 1 show that once the proposed system reaches sizes on the order of 100 MW, the CPE can be less than the targets outlined in previous studies, namely < $20/kWh, which are needed to reach full decarbonization of the grid.

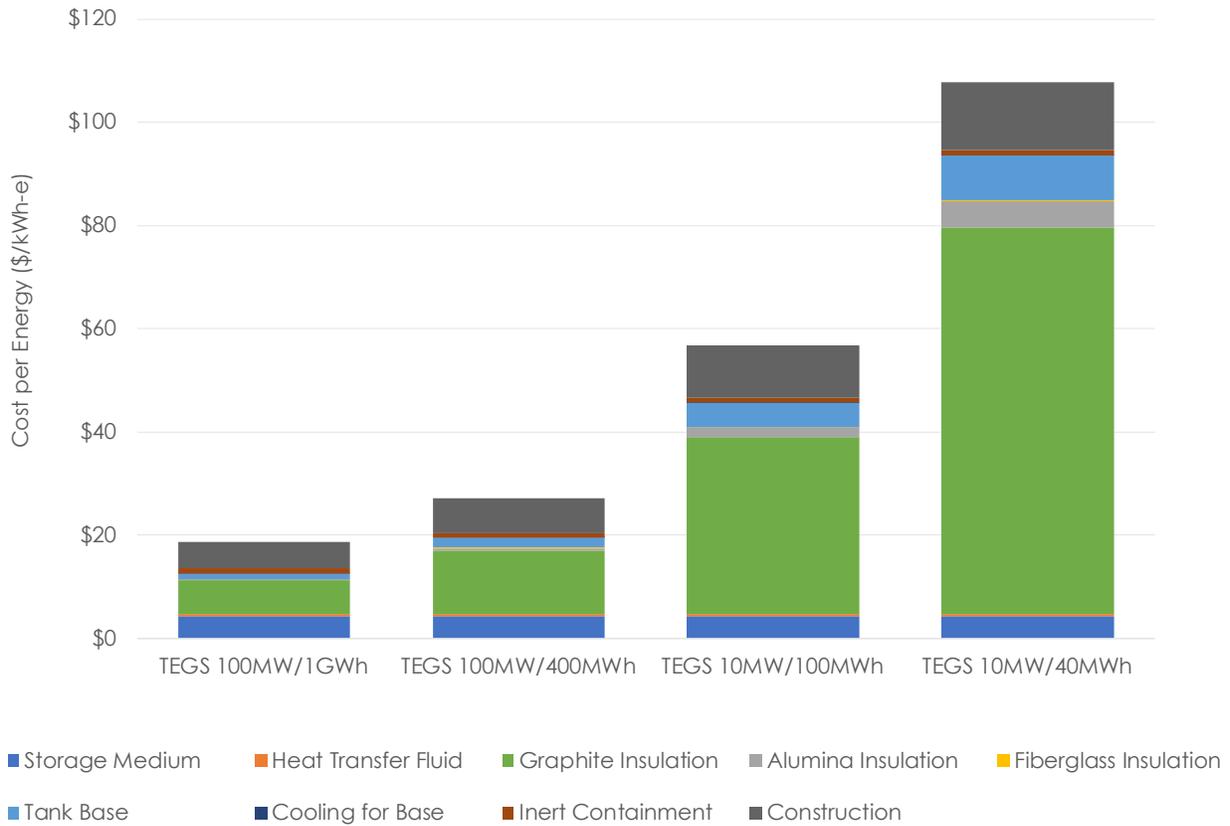

**Figure 2:** Cost per Energy at multiple scales for the energy storage component of TEGS

Figure 3 shows that in this new embodiment the CPP is unchanged, while Fig. 4 shows how the CAPEX, K, is affected by the changes with system size for CPE and the size independent CPP. These results indicate that the proposed system is one of the lowest cost options reported in the literature, and therefore increased attention and development is warranted. Based on our analysis, the CPE for this embodiment can be < $20/kWh-e, while the CPP can be < $0.40/W-e, which is considerably lower than the cost of alternative heat engine technologies such as a turbine. With these cost estimates, the system costs can reach ~$50/kWh-e as system sizes > 1 GWh are employed, which is lower than any other technology option we are aware of.

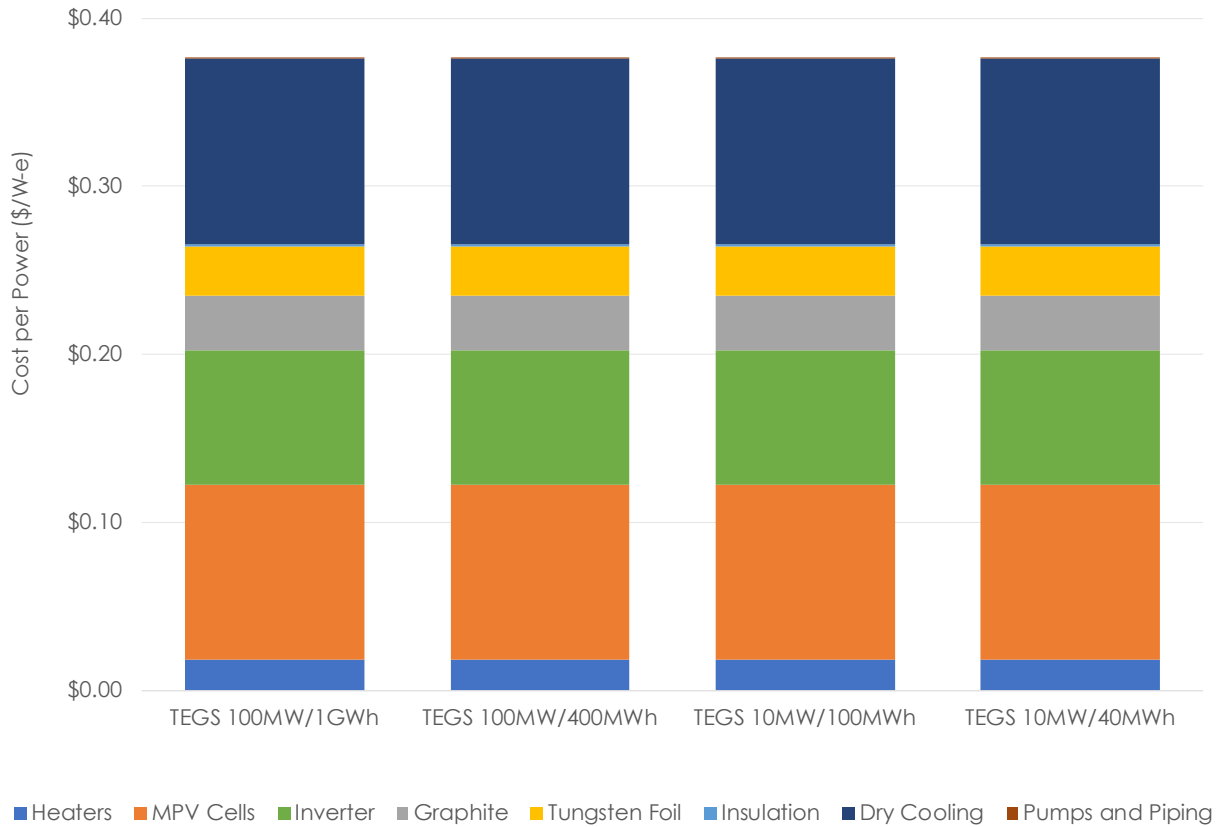

**Figure 3:** CPP for the power block at multiple scales. Due to the decoupled power and energy components of the TEGS system, this cost does not scale with size of the storage system and changes minimally with varying power scales.

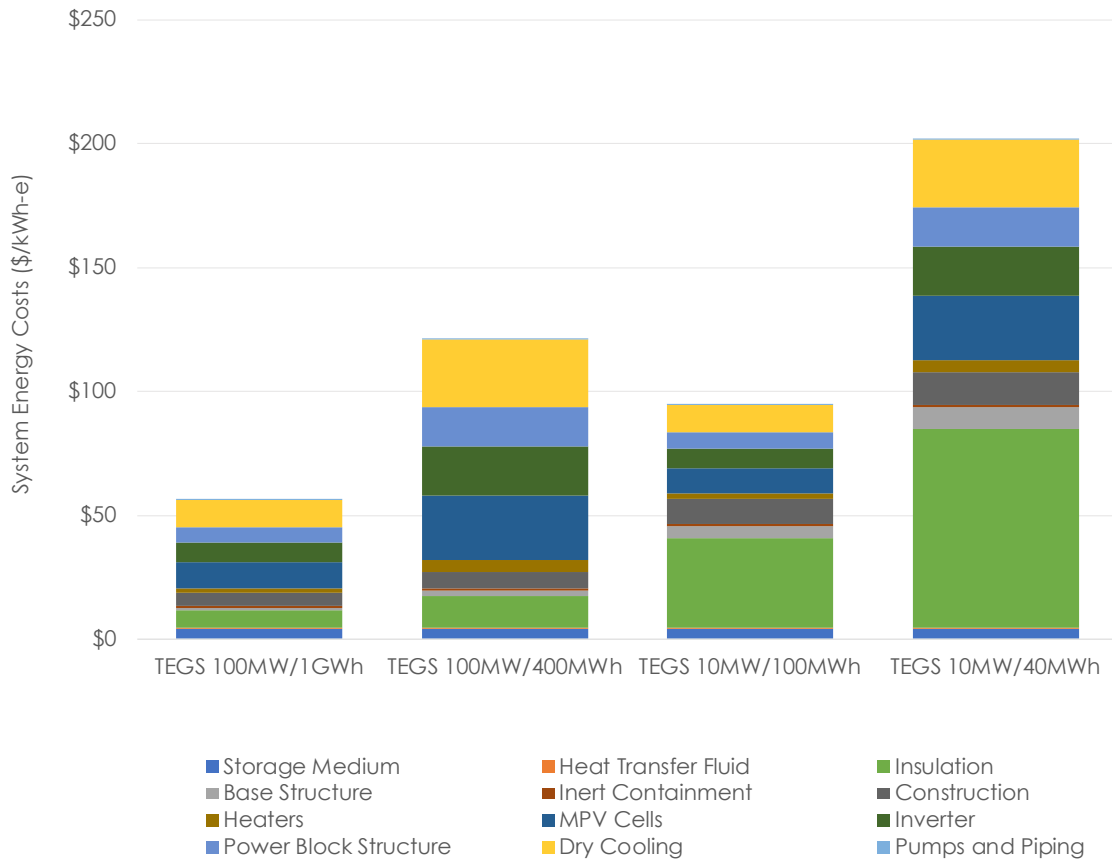

**Figure 4:** Total cost of energy incorporating the Cost per Energy and Cost per Power into a system energy cost for the four proposed cases.

**References:**

1. Albertus, P., Manser, J. S. & Litzelman, S. Long-Duration Electricity Storage Applications, Economics, and Technologies. *Joule* **4**, 21–32 (2020).
2. Sepulveda, N. A., Jenkins, J. D., Edington, A., Mallapragada, D. S. & Lester, R. K. The design space for long-duration energy storage in decarbonized power systems. *Nat. Energy* **6**, 506–516 (2021).
3. Ziegler, M. S. *et al.* Storage Requirements and Costs of Shaping Renewable Energy Toward Grid Decarbonization. *Joule* **3**, 2134–2153 (2019).
4. Amy, C., Seyf, H. R., Steiner, M. A., Friedman, D. J. & Henry, A. Thermal energy grid storage using multi-junction photovoltaics. *Energy Environ. Sci.* **12**, 334–343 (2019).
5. Amy, C. *et al.* Pumping liquid metal at high temperatures up to 1,673 kelvin. *Nature* **550**, 199–203 (2017).
6. Amy, C. Thermal Energy Grid Storage: Liquid Containment and Pumping. (Massachusetts Institute of Technology, 2020).